\documentclass[12pt,manuscript]{aastex}
\usepackage{natbib}
\usepackage{amsmath}
\usepackage{graphicx}

\shorttitle{Lower Bounds on IGMFs from Blazar GeV-TeV Light Curves}
\shortauthors{Takahashi et al.}

\begin{document}

\title{Lower Bounds on Intergalactic Magnetic Fields from
Simultaneously Observed GeV-TeV Light Curves of the Blazar Mrk 501}

\author{Keitaro Takahashi\altaffilmark{1,3,5},
        Masaki Mori\altaffilmark{2},
        Kiyotomo Ichiki\altaffilmark{3},
        Susumu Inoue\altaffilmark{4}
}

\altaffiltext{1}{Graduate School of Science and Technology,
Kumamoto University, 2-39-1 Kurokami, Kumamoto 860-8555, Japan}
\altaffiltext{2}{Department of Physical Sciences, Ritsumeikan University,
1-1-1 Noji Higashi, Kusatsu, Shiga 525-8577, Japan}
\altaffiltext{3}{Department of Physics and Astrophysics, Nagoya University,
Furo-cho, Chikusa-ku, Nagoya 464-8602, Japan}
\altaffiltext{4}{Institute for Cosmic Ray Research,
University of Tokyo, 5-1-5 Kashiwa-no-Ha, Kashiwa,
Chiba 277-8582, Japan}
\altaffiltext{5}{keitaro@sci.kumamoto-u.ac.jp}

\begin{abstract}
We derive lower bounds on intergalactic magnetic fields (IGMFs)
from upper limits on the pair echo emission from the blazar Mrk 501,
that is, delayed GeV emission from secondary $e^{-}e^{+}$ pairs produced
via interactions of primary TeV gamma rays with the cosmic infrared background.
Utilizing only simultaneous GeV-TeV light curves observed by VERITAS, MAGIC and {\it Fermi}-LAT
during a multiwavelength campaign in 2009 that included a TeV flare,
bounds are deduced on the IGMF strength
of $B \gtrsim 10^{-20}~{\rm G}$ at $90\%$ confidence level
for a field coherence length of 1 kpc.
Since our analysis is based firmly on the observational data alone and
nearly free of assumptions concerning the primary TeV flux
in unobserved periods or spectral bands,
our evaluation of the pair echo flux is conservative and
the evidence for a non-zero IGMF is
more robust compared to previous studies.
\end{abstract}

\keywords{BL Lacertae objects: individual (Mrk 501) --- gamma rays: observations ---
gamma rays: theory --- magnetic fields --- radiation mechanisms: nonthermal}

%%%%%%%%%%%%%%%%%%%%%%%%%%%%%%%%%%%%%%%%%%%%%%%%%%%%%%%%%%%%%%
\section{Introduction}
%%%%%%%%%%%%%%%%%%%%%%%%%%%%%%%%%%%%%%%%%%%%%%%%%%%%%%%%%%%%%%

Although their existence is yet to be observationally confirmed,
the possibility that weak magnetic fields were generated ubiquitously
in the early Universe has attracted considerable attention.
Such ``cosmological'' magnetic fields are potentially important from at least two perspectives. 
First, by serving as the seed fields for subsequent amplification by galactic dynamo mechanisms,
they may have been the ultimate origin of the magnetic fields seen today
in galaxies and clusters of galaxies (Widrow 2002).
Second, in some regions such as the centers of intergalactic voids,
they may have survived to the present day as intergalactic magnetic fields (IGMFs)
without being affected by later magnetization from astrophysical sources (Bertone et al. 2006),
and therefore may provide us with valuable, fossil information
about physical processes in the early Universe.
So far, various mechanisms have been proposed for the generation of such cosmological magnetic fields,
with predicted field amplitudes in the range $10^{-25}-10^{-15}~{\rm G}$
(e.g. Gnedin et al. 2000; Langer et al. 2005; Takahashi et al. 2005; Ichiki et al. 2006; Ando et al. 2010).
Although possibly sufficient as seeds for galactic dynamos,
such tiny magnetic fields are extremely difficult to confirm observationally
through conventional methods such as Faraday rotation measurements
or their effect on cosmic microwave background anisotropies (Widrow 2002).

In this context,
a potentially powerful probe of weak IGMFs with strengths $10^{-20}-10^{-15}~{\rm G}$
may be offered by pair echo emission from extragalactic TeV sources such as blazars or gamma-ray bursts
(e.g. Plaga 1995; Dai et al. 2002; Razzaque et al. 2004; Ichiki et al. 2008; Murase et al. 2008;
Takahashi et al. 2008, 2011; Neronov \& Semikoz 2009).
Pair echos comprise inverse Compton (IC) emission from secondary $e^{-}e^{+}$ pairs
produced via intergalactic $\gamma\gamma$ interactions among primary TeV gamma-rays
and infrared-UV photons of the extragalactic background light.
They are distinguishable through their characteristic time delay and spectral variation
that depend on the properties of the intervening IGMFs, while being insensitive
to galactic-scale magnetic fields either local to the source or the observer.
An alternative approach focusing on the spatial extension of the secondary gamma-ray halo emission
has also been discussed (Neronov \& Semikoz 2009; Ando \& Kusenko 2010).

Recently, Neronov \& Vovk (2010) claimed a lower bound on the IGMF of order $10^{-15}~{\rm G}$
from the non-detection of pair-echo components in the GeV spectra of selected TeV blazars.
A few other studies also gave similar results (Dolag et al. 2011; Tavecchio et al. 2010, 2011).
However, as pointed out by Dermer et al. (2011),
an implicit but crucial assumption in deriving these bounds
was that the TeV emission has been persistent for at least the past $10^6$ years,
at the level observed in these objects
on a small number of specific dates in the last several years.
This assumption is quite questionable,
because TeV blazars are generally known to be highly variable,
with their TeV flux fluctuating by more than two orders of magnitude
over timescales of several years and less,
for sufficiently well observed objects (e.g. Bartoli et al. 2011 and references therein).
With a more relaxed assumption that TeV gamma-rays from the blazar 1ES 0229+200
was steady only during the years 2005 to 2009 when the observations were conducted,
Dermer et al. (2011) obtained a much weaker lower bound of order $10^{-18}~{\rm G}$.
In fact, even this assumption is not entirely satisfactory, since
the actual time coverage of TeV observations in this period was very sparse,
being only several days in 2005, 2006 and 2009.
Taylor et al. (2011) have presented a similar analysis for a few blazars
with analogous assumptions on their TeV activity during unobserved periods.

Here we constrain the pair echo component from Mrk 501 and derive lower bounds on the IGMF,
relying solely on observational data from a multiwavelength campaign over a few months in 2009.
Thanks to a flare detected by VERITAS during this period,
lower bounds of order $10^{-20}~{\rm G}$ can be derived,
as the expected pair echo flux for weaker IGMFs would
exceed the concurrent upper limits on the daily GeV flux from {\it Fermi}-LAT.
Since we do not impose any assumptions on the TeV emission preceding the campaign,
our bounds implying non-zero IGMFs are much more robust than those obtained previously.

%%%%%%%%%%%%%%%%%%%%%%%%%%%%%%%%%%%%%%%%%%%%%%%%%%%%%%%%%%%%%%
\section{Mrk 501 observations \label{section:Mkn501}}
%%%%%%%%%%%%%%%%%%%%%%%%%%%%%%%%%%%%%%%%%%%%%%%%%%%%%%%%%%%%%%

Abdo et al. (2011) reported on a 4.5 month-long multiwavelength observing campaign of Mrk 501 in 2009,
featuring TeV data from VERITAS and MAGIC, and GeV data from {\it Fermi}-LAT.
The TeV spectra and light curves are essential input for evaluating the expected properties of the pair echo,
while the GeV data constrains the echo itself that typically emerges in this energy range (\S 3).

The VERITAS observing runs were conducted intermittently over a three month period MJD 54907-55004,
resulting in a total of 9.7 hours of good quality data.
During the 3-day ``flare state'' of MJD 54953-54955, the TeV flux increased significantly,
by about a factor of five relative to the remaining period, referred to as the ``quiescent state''.
The measured spectra for either state can be fitted by power-law functions of the form
$\log{F(E)} = \log{K} - a \log{(E/{\rm TeV})}$,
with $K = (4.17 \pm 0.24) \times 10^{-11} {\rm ph/cm^2 /sec /TeV}$
and $a = 2.26 \pm 0.06$ for the flare state, and
$K = (0.88 \pm 0.06) \times 10^{-11} {\rm ph/cm^2 /sec /TeV}$
and $a = 2.48 \pm 0.07$ for the quiescent state.
MAGIC was not available during the flare state but
gave a spectrum consistent with VERITAS for the quiescent state.
Both states are taken into account for calculating the pair echo emission.

On the other hand, {\it Fermi}-LAT performed uninterrupted monitoring of Mrk 501
in the survey mode during MJD 54683-55162, which included the whole campaign.
Although the GeV data did not reveal any strong flaring activity,
some mild flux variations were detected on timescales of about 30 days.
In contrast, changes in the contemporaneous GeV spectra were conspicuous, particularly
the 30-day spectrum for MJD 54952-54982 that was much harder than other periods.
Note that the first three days of this span correspond to the TeV flare state,
which may have possibly lasted much longer but was missed by
the sparse and irregular time coverage of TeV observations.
Nevertheless, below we conservatively assume the flare duration to be just these three days.

The Fermi-LAT data was acquired via the Fermi Science Support Center (FSSC)
\footnote{http://fermi.gsfc.nasa.gov/ssc/} and analyzed
to search for day-timescale flux variations around and after the flare.
We use the standard analysis tools supplied by FSSC in three energy bands,
100~MeV-1~GeV, 1~GeV-10~GeV and $>10$~GeV,
so as to keep reasonable photon statistics in time intervals as short as one day
while retaining some energy resolution.
Since the statistics is still small, we adopted the aperture photometry method,
i.e., events falling within one degree from the source were counted.
Note that background events above 1 GeV in one-day bins
at the high Galactic latitude of Mrk 501 are essentially negligible.
Flux upper bounds were calculated assuming Poisson statistics
when the data show no significant gamma-ray signal.

%%%%%%%%%%%%%%%%%%%%%%%%%%%%%%%%%%%%%%%%%%%%%%%%%%%%%%%%%%%%%%
\section{Pair echos and application to Mrk 501 \label{section:pair-echo}}
%%%%%%%%%%%%%%%%%%%%%%%%%%%%%%%%%%%%%%%%%%%%%%%%%%%%%%%%%%%%%%

First we briefly summarize the basic physics of pair echo emission
(see e.g. Ichiki et al. 2008 for more details),
and also provide an improved formulation for application to Mrk 501.
Primary gamma-rays with energy $E_{\gamma} \gtrsim 1 ~ {\rm TeV}$
emitted from an extragalactic source have mean free path
$
\lambda_{\gamma \gamma}
= 1/(0.26 \sigma_T n_{\rm IR})
= 190~{\rm Mpc}~( n_{\rm IR}/0.01 ~ {\rm cm}^{-3} )^{-1}
$
for $\gamma\gamma$ pair production interactions
with photons of the cosmic infrared background (CIB),
where $\sigma_T$ is the Thomson cross section and $n_{\rm IR}$
is the number density of CIB photons most relevant for the interactions.
The produced pairs with energy $E_e \approx E_{\gamma}/2$ 
give rise to the pair echo emission by IC upscattering of ambient cosmic microwave background (CMB) photons
to average energy
$
\langle E_{\rm echo} \rangle
= 2.7 T_{\rm CMB} \gamma_e^2
= 2.5~{\rm GeV}~
  ( E_{\gamma}/2 ~ {\rm TeV} )^2,
$
where $\gamma_e = E_e/m_e c^2$ is the Lorentz factor of the pairs
and $T_{\rm CMB} = 2.7~{\rm K}$ is the CMB temperature.
Thus, primary gamma-rays in the range $E_\gamma \simeq 1- 5~{\rm TeV}$
induce echos with typical energies $E_{\rm echo} \simeq 1-10~{\rm GeV}$.
The IC mean free path of the pairs is
$
\lambda_{\rm IC, scat}
= 1/(\sigma_T n_{\rm CMB}) = 1.2~{\rm kpc},
$
where $n_{\rm CMB} \approx 420~{\rm cm}^{-3}$ is the CMB photon
number density. The pairs upscatter CMB photons successively
until they lose most of their energy after propagating
an IC cooling length
$
\lambda_{\rm IC, cool}
= 3 m_e^2/(4 E_e \sigma_T U_{\rm CMB})
= 350 ~ {\rm kpc}~( E_e/1 ~ {\rm TeV} )^{-1},
$
where $U_{\rm CMB}$ is the CMB energy density.
The length scales for $\lambda_{\gamma \gamma}$ and $\lambda_{\rm IC, cool}$
imply that the secondary pairs typically arise in locations far removed from the source on scales of intergalactic voids,
whereas the pairs propagate only for short distances within such regions while generating the echo emission.

The pair echo emission arrives at the observer with a time delay relative to the primary emission,
caused by the effects of angular spreading in pair production and IC interactions,
as well as by deflections of the pairs in intervening magnetic fields.
The typical delay time due to angular spreading is
$
\Delta t_{\rm ang}
= (\lambda_{\gamma\gamma} + \lambda_{\rm IC, cool})/2 \gamma_e^2
\approx 3 \times 10^3 ~ {\rm sec} ~ (E_{\rm echo}/1 ~ {\rm GeV})^{-1} (n_{\rm IR}/0.01 ~ {\rm cm}^{-3})^{-1}
$
(Ichiki et al. 2008),
while that due to magnetic deflections is
$
\Delta t_{\rm B}
= (\lambda_{\gamma \gamma} + \lambda_{\rm IC, cool}) \langle \theta_{\rm B}^2 \rangle /2
$,
where
$
\langle \theta_{\rm B}^2 \rangle^{1/2}
= \max [\lambda_{\rm IC, cool}/r_{\rm L}, (\lambda_{\rm IC, cool} r_{\rm coh}/6)^{1/2} / r_{\rm L}]
$
is the variance of the magnetic deflection angle in terms of the Larmor radius $r_{\rm L}$
and field coherence length $r_{\rm coh}$.
If $r_{\rm coh} \ll \lambda_{\rm IC, cool}$,
$
\Delta t_{\rm B}
\approx 2 \times 10^4 ~ {\rm sec} ~  (E_{\rm echo}/1 ~ {\rm GeV})^{-3/2}  (B / 10^{-19} ~ {\rm G})^2
  (r_{\rm coh}/1 ~ {\rm kpc}) (n_{\rm IR}/0.01 ~ {\rm cm}^{-3})^{-1},
$
where $B$ is the field amplitude.
Hereafter we fiducially take $r_{\rm coh} = 1~{\rm kpc}$ (see e.g. Langer et al. 2005),
but the results are scalable to other values of $r_{\rm coh}$
since it enters only through the combination $B^2 r_{\rm coh}$
and only when $r_{\rm coh} \lesssim \lambda_{\rm IC, cool}$.
The total delay time is approximately
$\Delta t = \Delta t_{\rm ang} + \Delta t_{\rm B}$,
and the magnetic field properties are reflected in the delay
as long as $\Delta t_{\rm ang} \lesssim \Delta t_{\rm B}$.

The spectra and light curves of the pair echo can be evaluated as follows.
For a primary fluence $dN_{\gamma}/dE_{\gamma}$,
the associated time-integrated flux of secondary pairs is
\begin{equation}
\frac{dN_{e,{\rm 0}}}{d\gamma_e} (\gamma_e)
= 4 m_e
  \frac{dN_{\gamma}}{dE_{\rm \gamma}}(E_{\gamma} = 2 m_e \gamma_e)
  \left[1-e^{-\tau_{\gamma \gamma}(E_{\gamma} = 2 \gamma_e m_e)}\right],
\label{eq:dN0dgamma}
\end{equation}
where $\tau_{\gamma \gamma}(E_\gamma)$ is
the $\gamma\gamma$ optical depth in the CIB.
The time-dependent echo spectrum is
\begin{equation}
\frac{d^2 N_{\rm echo}}{dt dE_\gamma}
= \int d\gamma_e \frac{dN_e}{d{\gamma_e}}
  \frac{d^2 N_{\rm IC}}{dt dE_\gamma},
\end{equation}
where $d^2 N_{\rm IC}/dt dE_\gamma$ is the IC power from
a single electron or positron, and $dN_e/d{\gamma_e}$ is the total
time-integrated flux of pairs responsible for the echo emission
observed at time $t$,
related nontrivially to $dN_{e, \rm 0}/d{\gamma_e}$ in Eq. (\ref{eq:dN0dgamma}).
If the distance to the source from the observer $D \gg \lambda_{\gamma \gamma}$,
it can be approximated by
$dN_e/d{\gamma_e} =
 (\lambda_{\rm IC, cool}/c \Delta t) dN_{e,{\rm 0}}/d\gamma_e$ (Dai et al. 2002),
but this is not the case for Mrk 501 whose $D \sim 130$ Mpc
can be comparable to $\lambda_{\gamma \gamma}$ for $E_\gamma \gtrsim$ TeV.
In order to improve the evaluation of $dN_e/d{\gamma_e}$
by accounting for the finite probability of pair production near the observer,
we consider the time-integrated flux due to the fraction of pairs that originate
between radii $r_{\gamma\gamma}$ and $r_{\gamma\gamma} + \Delta r$ from the source,
\begin{equation}
\Delta \frac{dN_e}{d\gamma_e}
= 4 m_e \frac{dN_{\gamma}}{dE_{\gamma}}
  \left( e^{-r_{\gamma\gamma}/\lambda_{\gamma\gamma}}
         - e^{-(r_{\gamma\gamma} + \Delta r)/\lambda_{\gamma\gamma}}
  \right).
\end{equation}
The total time-integrated flux of pairs can be evaluated
by integrating over $r_{\gamma\gamma}$ as
\begin{equation}
\frac{dN_e}{d\gamma_e}
= \int_{0}^{D} dr_{\gamma\gamma} \frac{\lambda_{\rm IC, cool}}{c \Delta t (r_{\gamma\gamma})}
                   4 m_e \frac{dN_{\gamma}}{dE_{\gamma}}
                   \frac{e^{-r_{\gamma\gamma}}}{\lambda_{\gamma\gamma}},
\end{equation}
where $\Delta t (r_{\gamma\gamma})$
is given by the above expression for $\Delta t = \Delta t_{\rm ang} + \Delta t_{\rm B}$
with $\lambda_{\gamma\gamma}$ replaced by $r_{\rm \gamma\gamma}$.

Note that the pair echo fluence
is determined by the total amount of absorbed primary gamma rays
and thus independent of the IGMF,
in contrast to the pair-echo flux which is roughly given by the fluence divided by $\Delta t$.
Weaker IGMFs generally give higher echo fluxes,
as long as the time delay does not become dominated by angular spreading
and the echo flux remains sensitive to $B$.
For $r_{\rm coh} = 1~{\rm kpc}$,
$\Delta t_{\rm B}$ approaches $\Delta t_{\rm ang}$ if $B \sim 10^{-20}~{\rm G}$.

In applying the above formalism to the 2009 Mrk 501 activity,
we clarify what we employ for the primary TeV spectra and light curves.
Both flare and quiescent states can make important contributions to the pair-echo emission.
A TeV flare was observed for at least three days from MJD 54953 (\S 2);
however, it may have continued for a longer time,
or even separate flares could have occurred over the following weeks,
as can be speculated from the hard, 30-day {\it Fermi}-LAT spectrum.
Nevertheless, we choose to be conservative and assume that
there is no other flare state during the campaign besides the three days seen by VERITAS.
Although the quiescent state was also only sparsely sampled at TeV,
since both VERITAS and MAGIC measured a consistent flux and spectrum at separate times,
we postulate that the quiescent emission is steady over the period covered by the TeV telescopes,
the sole assumption we make regarding TeV activity not directly observed.
Thus, we consider the primary light curve to consist of a flare state with a top-hat shape
for the 3 days MJD 54953-54955, together with a steady quiescent state
for the preceding 46 days MJD 54907-54952 as well as the ensuing 49 days MJD 54956-55004.
The primary flux and spectrum for each state are chosen such that
they are compatible with those observed by VERITAS
after accounting for intergalactic $\gamma\gamma$ absorption
with the CIB model of Franceschini et al. (2008),
and are described by the same power-law functional form mentioned in \S 2
but with the parameters
$K = 9 \times 10^{-11} {\rm ph/cm^2 /sec /TeV}$ and $a = 2.0$ for the flare state, and
$K = 2 \times 10^{-11} {\rm ph/cm^2 /sec /TeV}$ and $a = 2.3$ for the quiescent state.
Minimum and maximum cutoffs are also imposed at 0.1 TeV and 5 TeV, respectively,
the latter corresponding to the highest energy photons detected by VERITAS and MAGIC.
Comparing the pair echo emission calculated in this way with
the observed GeV limits gives conservative lower bounds on the IGMF,
since any additional primary emission, outside either the above time interval or the above spectral range,
would only add to the pair echo flux and lead to tighter bounds.

%%%%%%%%%%%%%%%%%%%%%%%%%%%%%%%%%%%%%%%%%%%%%%%%%%%%%%%%%%%%%%
\section{Results and discussion \label{section:results}}
%%%%%%%%%%%%%%%%%%%%%%%%%%%%%%%%%%%%%%%%%%%%%%%%%%%%%%%%%%%%%%

\begin{figure}[!htb]
\epsscale{1}
\plotone{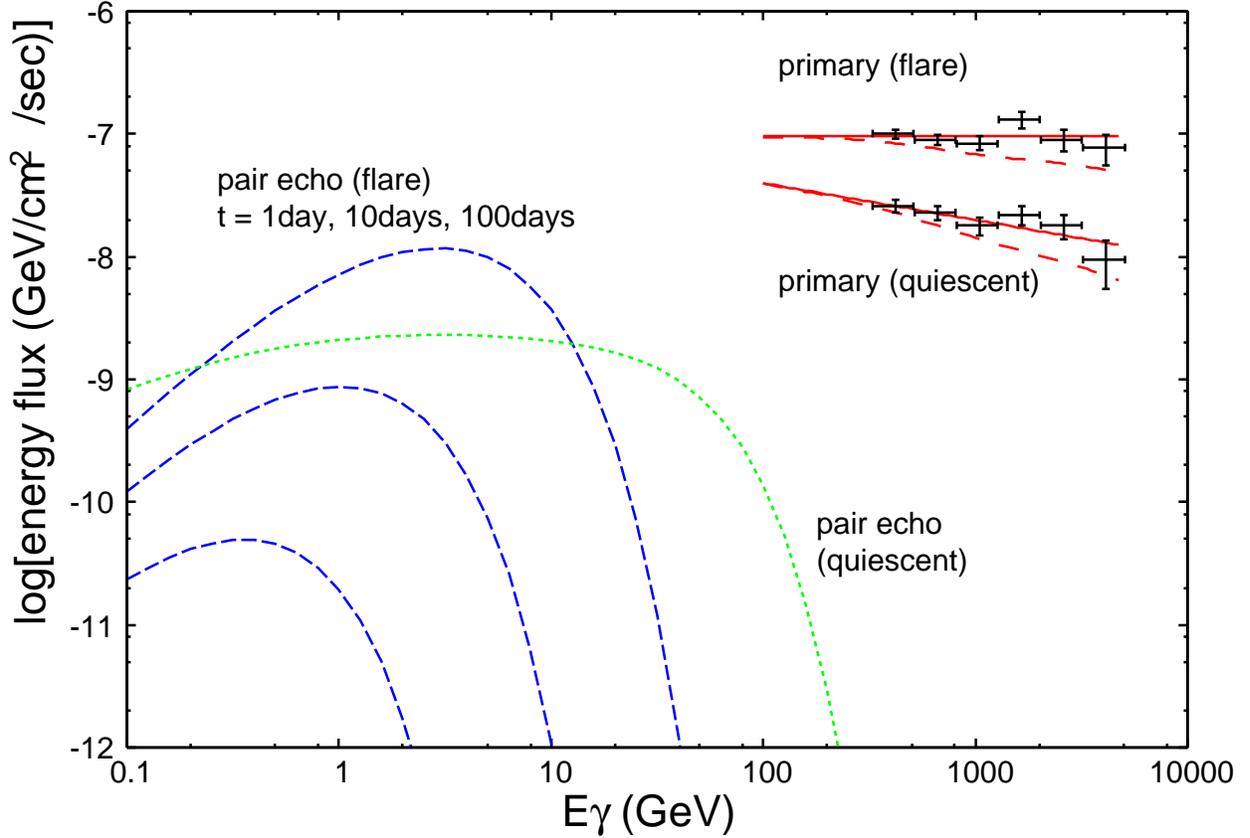}
\caption{Spectra of the primary and pair echo emission of Mrk 501 for the case $B = 10^{-20}~{\rm G}$.
The primary spectra for the flare and quiescent states are each shown
with (long-dashed)
and without (solid)
intergalactic $\gamma\gamma$ absorption,
along with the absorption-corrected data from VERITAS observations.
Also plotted are the echo from the flare state at observer time
$t=$1, 10 and 100 days after the flare
(dashed, from top to bottom),
as well as the echo from the quiescent state (dotted).}
\label{fig:spectrum}
\end{figure}

Fig. \ref{fig:spectrum} shows the spectra of the primary and pair echo emission
for the flare and quiescent states when $B = 10^{-20}~{\rm G}$.
The primary spectra are displayed both with and without intergalactic $\gamma\gamma$ absorption,
the latter to be compared with the absorption-corrected VERITAS data as given in Abdo et al. (2011).
The echo from the flare state is plotted at observer times
$t=$1, 10 and 100 days after the flare,
fading progressively on timescales approximately corresponding to $\Delta t$.
In contrast, here the echo due to the quiescent state is essentially stationary
on the timescale of the campaign and independent of $B$.
Note, however, that for stronger $B$ with accordingly longer $\Delta t_B$,
even the quiescent echo component can be nonstationary, particularly at low energies.
Only when the primary emission persists at a steady level for a time
considerably longer than $\Delta t_B$
does the echo reach stationarity, as demonstrated by Dermer et al (2011).

\begin{figure}[!htb]
\epsscale{1}
\plotone{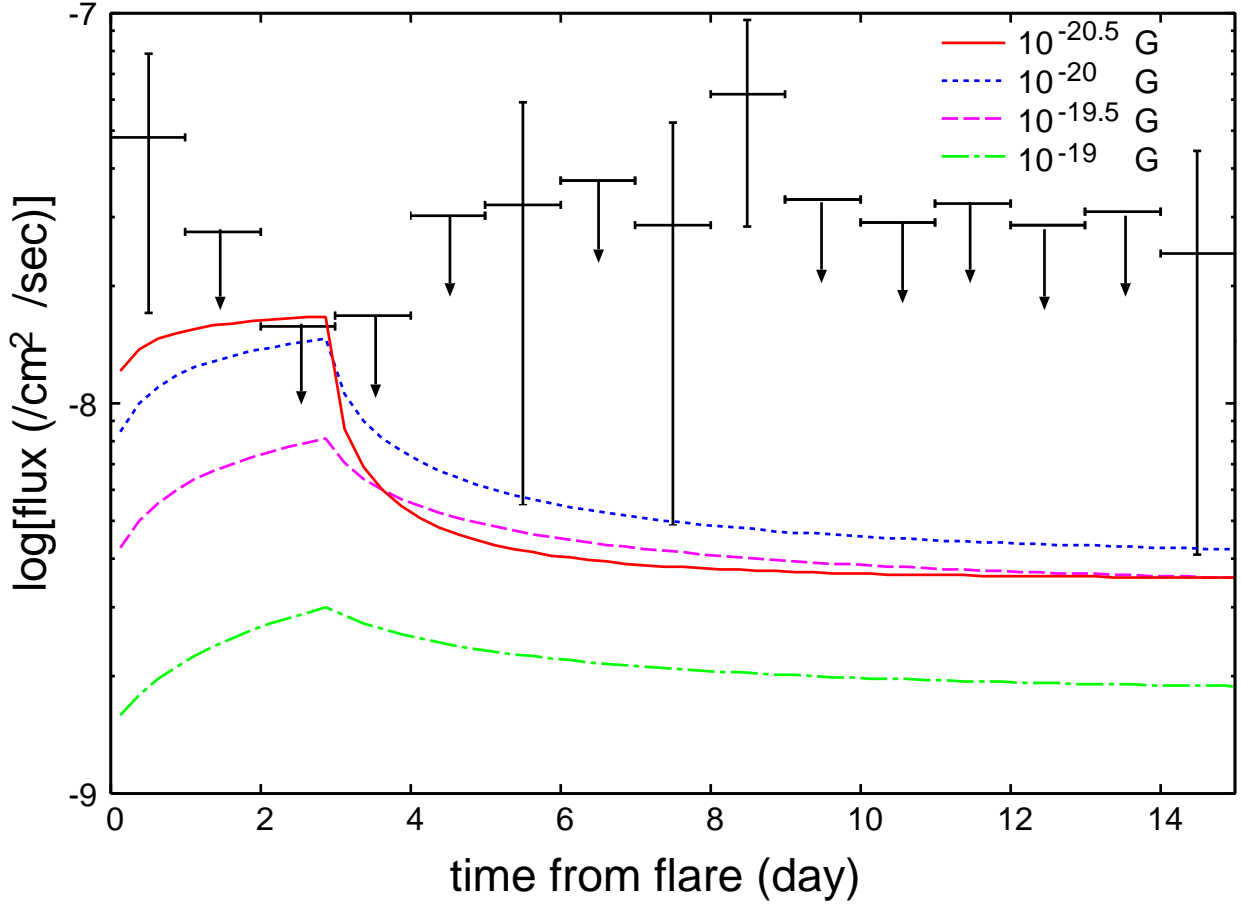}
\caption{Light curve of Mrk 501 in the 1-10 GeV band from the onset of the TeV flare on MJD 54953.
Pair echo expectations for $B = 10^{-20.5}, 10^{-20}, 10^{-19.5}, 10^{-19}~{\rm G}$ (curves from top to bottom)
are compared with {\it Fermi}-LAT data binned at 1-day intervals using the aperture photometry method,
where errors (vertical bars) or upper limits (downward arrows) are at $68 \%$ confidence level.}
\label{fig:lc}
\end{figure}

The light curves of the pair echo in the 1-10 GeV band after the onset of the TeV flare on MJD 54953
are plotted in Fig. \ref{fig:lc} for different values of $B$.
The flux initially rises over the duration of the primary flare,
and then decays roughly exponentially on timescales of order $\Delta t$
as the primary emission switches to the quiescent state.
At sufficiently late times, only the quiescent emission contributes to the echo
and its flux approaches a steady level.
For weaker $B$, the echo flux responds to changes of the primary flux
more quickly by virtue of the shorter $\Delta t_B$,
and the peak flux is larger and more susceptible to observational limits.

Thus, in order to observationally constrain IGMFs in the range $B \sim 10^{-20}-10^{-19}~{\rm G}$,
GeV-band light curves with time resolution of order a day are necessary.
We analyzed the {\it Fermi}-LAT data to obtain 1-10 GeV gamma-ray fluxes and upper limits
with 1-day time binning, the results of which are compared with the pair echo expectations in Fig. \ref{fig:lc}.
Most are upper limits, which is not surprising given the limited sensitivity of {\it Fermi}-LAT
with such short integration times.
The strongest limits come from the second, third and fourth days,
being comparable to the pair-echo predictions for $B \lesssim 10^{-20}~{\rm G}$
and providing important lower bounds on IGMFs in this range,
while higher values of $B$ cannot be usefully constrained by the current
analysis. Note that primary emission in the GeV range is also generally
expected and should contribute significantly to the LAT light curve,
which implies that the actual pair echo flux is even lower and the true IGMF
lower bounds stronger. However, given the lack of reliable knowledge on
the primary GeV spectra and variability, we restrict ourselves to
conservative constraints by not accounting for any such components.

To deduce bounds on $B$, we first calculate from the obtained LAT upper
limits the probability for each day that the predicted pair-echo flux for
a specific value of B does not exceed the true flux. Then we regard each
value of $B$ as being allowed at the probability equal to the product of
the above daily probabilities. Thus we arrive at our main result
that $B \gtrsim 10^{-20}~{\rm G}$ at about $90\%$ confidence level,
determined mostly by the limits from the second through fourth days.
We have also carried out similar analyses for other energy bands
$<1$ GeV or $>10$ GeV but were not able to derive significant constraints,
as can be expected from Fig. \ref{fig:spectrum}.

In summary, by comparing the daily GeV flux upper limits from {\it Fermi}-LAT
for the blazar Mrk 501 during and after its TeV flare in 2009
with the expected light curves for the the associated pair echo emission,
we have derived lower bounds on IGMF strengths of
$B \gtrsim 10^{-20}~{\rm G}$ at $90\%$ confidence level
for a field coherence length $r_{\rm coh} = 1~{\rm kpc}$.
The result can be roughly scaled for other values of $r_{\rm coh}$ as
$B \gtrsim 5 \times 10^{-22}~\max[(r_{\rm coh}/350~{\rm kpc})^{-1/2},1]~{\rm G}$.
This bound is weaker compared to other recent results obtained through similar methods,
which, however, all relied on unproven assumptions regarding the TeV emission during unobserved periods
on timescales of years (Dermer et al. 2011; Taylor et al. 2011)
or much longer (Neronov \& Vovk 2010; Dolag et al. 2011; Tavecchio et al. 2010, 2011).
In constrast, our analysis is entirely free of such assumptions other than for the quiescent state,
and thus can be considered the most robust indication so far
for the existence of non-zero IGMFs.

Future progress can be expected from
regular, long-term coverage of the multi-TeV emission
by wide-field facilities such as HAWC or LHAASO,
as well as GeV-TeV spectral variability measurements
with high sensitivity and time resolution by CTA
to disentangle and positively identify the echo component
from the primary emission.
If IGMFs are somewhat stronger,
the spatially-extended pair halo emission may be detectable by CTA.
Such advances will surely pave a new road toward understanding cosmic magnetic fields.

\acknowledgments

We thank Andrii Neronov for crucial comments.
This work is supported by Grants-in-Aid from MEXT of Japan,
No.~21840028 (KT), No.~22540315 (MM), No.~21740177, No.~22012004 (KI) and
No.~22540278 (SI),
and for the global COE program
``The Next Generation of Physics''
at Kyoto University and
``Quest for Fundamental Principles in the Universe''
at Nagoya University.


\begin{thebibliography}{99}
\bibitem[Abdo et al. 2011]{Abdo}
Abdo, A. A., et al. 2011, ApJ, 727, 129
\bibitem[Ando et al. 2010]{AndoDoiSusa}
Ando, M., Doi, K., \& Susa, H. 2010, ApJ, 716, 1566
\bibitem[Ando \& Kusenko 2010]{AndoKusenko}
Ando, S., \& Kusenko, A. 2010, ApJ, 722, L39
\bibitem[Bartoli et al. 2011]{Bartoli}
Bartoli, B. et al. 2011, ApJ, 734, 110
\bibitem[Bertone et al. 2006]{Bertone}
Bertone, S., Vogt, C., \& En{\ss}lin, T. 2006, MNRAS, 370, 319
\bibitem[Dai et al. 2002]{Dai}
Dai, Z. G., Zhang, B., Gou, L. J., M{\'e}sz{\'a}ros, P., \& Waxmann, E.
2002, ApJ, 580, L7
\bibitem[Dermer et al. 2011]{Dermer}
Dermer, C. D., Cavadini, M., Razzaque, S., Finke, J. D., \& Lott, B.
2011, ApJ, 733, L21
\bibitem[Dolag et al. 2011]{Dolag2}
Dolag, K., Kachelrie{\ss}, M., Ostapchenko, S., \& Tom{\'a}s, R.
2011, ApJ, 727, L4
\bibitem[Franceschini et al. 2008]{Franceschini}
Franceschini, A., Rodighiero, G., \& Vaccari, M. 2008, A{\&}A, 487, 837
\bibitem[Gnedin et al. 2000]{Gnedin}
Gnedin, N. Y., Ferrara, A., \& Zweibel, E. G. 2000, ApJ, 539, 505
\bibitem[Ichiki et al. 2008]{Ichiki1}
Ichiki, K., Inoue, S., \& Takahashi, K. 2008, ApJ, 682, 127
\bibitem[Ichiki et al. 2006]{Ichiki2}
Ichiki, K., Takahashi, K., Ohno, H., Hanayama, H., \& Sugiyama, N.
2006, Science, 311, 827
\bibitem[Langer et al. 2005]{Langer}
Langer, M., Aghanim, N., \& Puget, J.-L. 2005, A{\&}A, 443, 367
\bibitem[Murase et al. 2008]{Murase2}
Murase, K., Takahashi, K., Inoue, S., Ichiki, K., \& Nagataki, S.
2008, ApJ, 686, L67
\bibitem[Neronov \& Semikoz 2009]{Neronov1}
Neronov, A., \& Semikoz, D. V. 2009, Phys. Rev. D, 80, 123012
\bibitem[Neronov \& Vovk 2010]{Neronov3}
Neronov, A., \& Vovk, I. 2010, Science, 328, 73
\bibitem[Plaga 1995]{Plaga}
Plaga, R. 1995, Nature, 374, 430
\bibitem[Razzaque et al. 2004]{Razzaque}
Razzaque, S., M{\'e}sz{\'a}ros, P., \& Zhang, B. 2004, ApJ, 613, 1072
\bibitem[Takahashi et al. 2005]{Takahashi1}
Takahashi, K., Ichiki, K., Ohno, H., \& Hanayama, H.
2005, Phys. Rev. Lett., 95, 121301
\bibitem[Takahashi et al. 2011]{Takahashi2}
Takahashi, K., Inoue, S., Ichiki, K., \& Nakamura, T. 2011, MNRAS, 410, 2741
\bibitem[Takahashi et al. 2008]{Takahashi3}
Takahashi, K., Murase, K., Ichiki, K., Inoue, S., \& Nagataki, S.
2008, ApJ, 687, L5
\bibitem[Tavecchio et al. 2010]{Tavecchio1}
Tavecchio, F., Ghisellini, G., Foschini, L., Bonnoli, G., Ghirlanda, G., \& Coppi, P.
2010, MNRAS, 406, L70
\bibitem[Tavecchio et al. 2011]{Tavecchio2}
Tavecchio, F., Ghisellini, G., Bonnoli, G., \& Foschini, L.
2011, MNRAS, 414, 3566
\bibitem[Taylor et al. 2011]{Taylor}
Taylor, A. M., Vovk, I., \& Neronov, A. 2011, A\&A, 529, A144
\bibitem[Widrow 2002]{Widrow}
Widrow, L. M. 2002, Reviews of Modern Physics, 74, 775
\end{thebibliography}
\end{document}